\documentclass[lettersize,journal]{IEEEtran}

\usepackage{amsmath,amsfonts}
\usepackage{array}
\usepackage[caption=false,font=normalsize,labelfont=sf,textfont=sf]{subfig}
\usepackage{textcomp}
\usepackage{stfloats}
\usepackage{url}
\usepackage{graphicx}
\usepackage{float}
\usepackage{cite}
\usepackage{hyperref}
\usepackage{xcolor}
\usepackage{eso-pic}
\definecolor{ariescolor}{HTML}{D13064}
\definecolor{aifscolor}{HTML}{3B82F6}
\definecolor{eccolor}{HTML}{10B981}
\graphicspath{{figures/}}

\newlength{\logolength}
\AddToShipoutPictureFG{%
  \AtPageLowerLeft{%
    \put(\LenToUnit{\dimexpr\paperwidth-0.40in},\LenToUnit{\dimexpr0.5in+\logolength}){%
      \rotatebox{-90}{\includegraphics[width=\logolength]{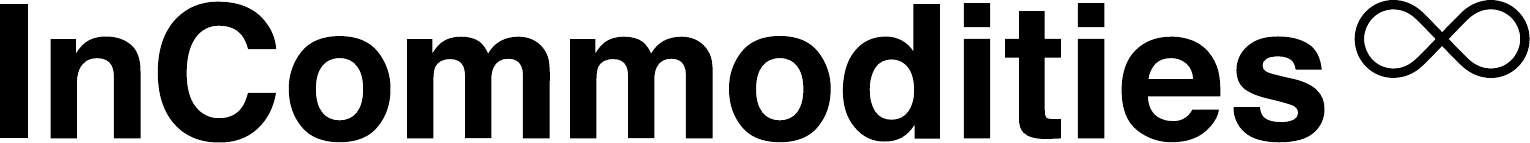}}%
    }%
  }%
}

\begin{document}

\title{Aries: A Proprietary Medium-Range Weather Prediction Model for the Energy Industry}

\author{Lukas Hedegaard Morsing, Arian Bakhtiarnia, Jonas Lynge Olesen, \\
   Tómas Bragi Björnsson Leth, and Christian Gøbel Bach \\
  InCommodities$^\infty$
  \thanks{$^\infty$https://incommodities.com}
  }

\maketitle

\begin{abstract}
    Medium-range weather forecasting underpins operational and planning decisions across the energy industry.
    Developing competitive weather models was once the domain of national meteorological centers, but recent advances in machine-learned weather prediction (MLWP) have opened the field to industry.
    We present Aries, a SwinTransformer-based MLWP model developed at InCommodities.
    Aries is trained on ERA5 reanalysis data at 0.25\textdegree{} resolution, predicting 74 prognostic and 11 diagnostic atmospheric variables.
    We evaluate the model on 2025 ECMWF Analysis initializations, ensuring a recent and strictly out-of-sample test period for all models compared.
    On 10-metre wind speed, Aries outperforms both ECMWF HRES and AIFS in terms of RMSE for lead times up to four days, while on 2-metre temperature it achieves RMSE on par with AIFS operational.
    These results demonstrate that proprietary development of competitive weather models is technically viable, supporting a broader set of forecasts available for operational and planning applications in the energy industry.
\end{abstract}

\section{Introduction}

Weather conditions determine renewable generation and electricity demand. Together with transmission constraints, they shape regional power balances and cross-border flows, ultimately driving power prices. Accurate weather forecasts from hours to weeks ahead are therefore critical infrastructure for energy markets, translating directly into improved price forecasts and more efficient market outcomes.
Renewables supplied 32\% of global electricity in 2024 \cite{iea2025globalreview}, with the European Union reaching 47.5\% \cite{eurostat2025renewables}, and as renewable penetration grows the sensitivity of the power system to forecast accuracy intensifies \cite{staffell2018weather}.
Short- and medium-range weather forecasts govern auction-based trading for next-day delivery and continuous intraday trading, where forecast errors translate directly into imbalances between supply and demand.

Numerical weather prediction (NWP) has served this role for decades, but the past three years have seen machine-learned weather prediction (MLWP) models emerge as a credible alternative.
Several systems now match or surpass the deterministic skill of leading NWP models at a fraction of the computational cost and latency \cite{lam2023graphcast, bodnar2024aurora, moldovan2025aifs}, and publicly available training data (ERA5 \cite{hersbach2020era5}) combined with commodity-GPU inference has lowered the barrier to entry well beyond national meteorological agencies.

Industry players have begun to follow suit, developing in-house MLWP systems alongside the established public-sector providers.
At InCommodities we developed Aries as an independently trained alternative.
We present evaluations on two central weather variables: 2\,m temperature, which strongly informs energy demand, and 10\,m wind speed, which drives wind turbine power generation.
We demonstrate deterministic skill that surpasses ECMWF HRES and AIFS on 10\,m wind speed at lead times up to four days and matches AIFS on 2\,m temperature.

\begin{figure}[t]
    \centering
    \includegraphics[width=\columnwidth,trim=0 30 0 0,clip]{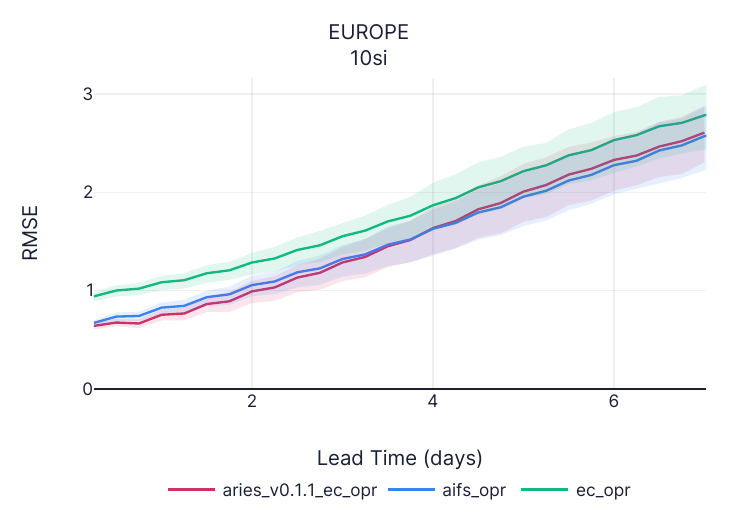}
    \caption{Latitude-weighted RMSE of 10\,m wind speed forecasts over Europe for operational 2025 runs with
    {\color{ariescolor}\textbf{---}}~Aries,\;
    {\color{aifscolor}\textbf{---}}~AIFS, and\;
    {\color{eccolor}\textbf{---}}~HRES against ERA5.
    Shaded regions denote 25th--75th percentile ranges.
    }
    \label{fig:rmse_10si_eu}
\end{figure}

\section{Background: From NWPs to MLWPs}
\label{sec:background}

NWP produces forecasts by solving discretized partial differential equations (PDEs) of atmospheric dynamics on a global grid.
ECMWF's Integrated Forecasting System (IFS), and in particular its high-resolution deterministic configuration (HRES), has long been the gold standard for medium-range forecasting, but each cycle requires hours of computation on dedicated supercomputers.

Since 2022, data-driven models trained on ERA5 reanalysis \cite{hersbach2020era5} have rapidly closed this gap.
FourCastNet \cite{pathak2022fourcastnet}, Pangu-Weather \cite{bi2023pangu}, GraphCast \cite{lam2023graphcast}, FuXi \cite{chen2023fuxi}, Aurora \cite{bodnar2024aurora}, and ECMWF's own AIFS \cite{moldovan2025aifs} now match or exceed HRES on most evaluation targets, while Jua.ai's EPT-2 \cite{molinaro2025ept2} demonstrates that startup players can produce competitive MLWPs.
Because inference requires only a single forward pass, a 10-day global forecast takes seconds on a commodity GPU, dramatically lowering the barrier to entry.

A shared limitation of deterministic MLWPs trained with MSE or MAE losses is spectral blurring at longer lead times: The loss penalizes sharp but misplaced features, incentivizing smooth predictions \cite{willard2024training, lam2023graphcast}.
Multi-step autoregressive fine-tuning improves RMSE but can exacerbate this effect \cite{willard2024training, moldovan2025aifs}.
Frequency-domain losses offer a partial remedy \cite{subich2025amse, chattopadhyay2023fourks}, while CRPS-based and generative approaches such as GenCast \cite{price2024gencast} and AIFS-CRPS \cite{lang2024aifscrps} mitigate the problem by incentivizing realistic ensemble spread rather than a smoothed mean.

\section{Model Architecture and Operational Setting}
\label{sec:architecture}

\subsection{Architecture Overview}

Aries is built on the Shifted Window Transformer architecture derived from
SwinV2~\cite{liu2022swinv2}.
The backbone is \emph{non-hierarchical}: All
layers operate at the same spatial feature resolution, avoiding the
down-sampling stages typical of vision transformers.
The input atmospheric state
is split into patches that serve as tokens; shifted-window self-attention is
then applied with periodic zonal wrap-around padding to respect the spherical topology of the globe.
Table~\ref{tab:model_size} summarizes the model configuration.

\begin{table}[h]
    \centering
    \caption{Aries model configuration.}
    \label{tab:model_size}
    \begin{tabular}{ll}
        \hline
        \textbf{Parameter}       & \textbf{Value} \\
        \hline
        Depth                    & 14 layers      \\
        Attention heads          & 16             \\
        Embedding dimension      & 3{,}584        \\
        Total parameters         & 2.4\,B         \\
        \hline
    \end{tabular}
\end{table}

\subsection{Formulation}
Aries ingests a \emph{single} atmospheric state and predicts the next state in one forward pass.
The model predicts a residual update that is added to the input state:
\begin{equation}
    \tilde{X}^{t+\Delta t} = f_\theta(X^t) + r(X^t)
\end{equation}
where $f_\theta$ denotes the learned update mapping, $r$ is a residual transformation, and $X^t$ is the atmospheric state at time $t$, and $\Delta t = 6$\,h.
We employ a modified residual connection $r$ inspired by low-pass filtering approaches in the literature~\cite{chattopadhyay2023fourks, kochkov2024neuralgcm}.




\subsection{Training}

The model is trained on ERA5 reanalysis data~\cite{hersbach2020era5} at
0.25\textdegree{} resolution ($720 \times 1440$ grid points).
The training period spans 1979--2022, with 2023 reserved for validation and 2024 for testing.
Training proceeds in two distinct phases:
(i)~\emph{single-step training}, in which the model learns to predict one 6\,h step ahead, and
(ii)~\emph{multi-step autoregressive fine-tuning}, in which the model is
unrolled over successively more steps and optimized end-to-end. We trained Aries up to 10 roll-out steps (60\,h lead time).

Training uses a latitude-weighted MSE loss for the
single-step phase, switching to a latitude-weighted L1 loss during
autoregressive fine-tuning.
In both phases, we use channel-wise weights that follow the scheme of
Lam et al.~\cite{lam2023graphcast}, with increased weight on wind components given their direct relevance to energy applications.
An auxiliary frequency-domain penalty term is applied at low
weight to mitigate spectral degradation at longer lead
times~\cite{chattopadhyay2023fourks}.
This term operates on the power spectral density of each channel:
\begin{equation}
    \mathcal{L}_{\text{freq}} = \frac{1}{C} \sum_{c=1}^{C} \sum_{\substack{k_x,\, k_y \,\geq\, k_{\min}}} \left| \frac{S_c(\tilde{X})_{k_x, k_y} - S_c(X)_{k_x, k_y}}{Z_c} \right|,
    \label{eq:spectral_loss}
\end{equation}
where $S_c(X) = |\mathcal{F}_c(X)|^2$ is the 2D power spectral density of channel $c$, $\mathcal{F}$ denotes the discrete Fourier transform over latitude and longitude, $Z_c = \bigl|\operatorname{mean}_{k_x, k_y} S_c(X)_{k_x, k_y}\bigr|$ normalizes per channel, and wavenumbers below $k_{\min}$ are excluded to focus penalization on fine-scale structure.



\section{Benchmark Setup}
\label{sec:benchmarks}

All models are evaluated over the full calendar year 2025, initialized from ECMWF operational analysis at the 00Z and 12Z cycles.
Forecasts are verified against ERA5 reanalysis \cite{hersbach2020era5}.
The year 2025 predictions stem from operational runs for all models.
We compare Aries against ECMWF HRES (the operational NWP gold standard) and AIFS \cite{moldovan2025aifs}.
Note that unlike HRES and AIFS, Aries was never fine-tuned with ECMWF operational analysis inputs.

Metrics are computed over Europe and the USA\footnote{Europe latitude [32.2, 72.0], longitude [-15.6, 36.4]. USA latitude [23.8, 52.2], longitude [-125.6, 58.4].}, latitude-weighted over all grid points.
We evaluate 10\,m wind speed (10si) and 2\,m temperature (2t).

Let $e_i(t) = \tilde{X}_i(t) - X_i(t)$ denote the forecast error at grid point $i$ and initialization time $t$, with latitude-weighted spatial mean
\begin{equation}
    \bar{e}(t) = \frac{\sum_{i} w_i \, e_i(t)}{\sum_i w_i}, \quad w_i = \cos(\text{lat}_i),
\end{equation}
which also corresponds to the \emph{Bias}.
The latitude-weighted root mean square error (\emph{RMSE}) is defined as
\begin{equation}
    \text{RMSE} = \sqrt{\frac{\sum_i w_i \, e_i^2}{\sum_i w_i}}.
\end{equation}

\section{Deterministic Forecasting Results}

\subsection{10m Wind Speed}

\begin{figure*}[!tb]
    \centering
    \begin{minipage}[t]{0.32\textwidth}
        \centering
        \includegraphics[width=\textwidth,trim=0 30 0 0,clip]{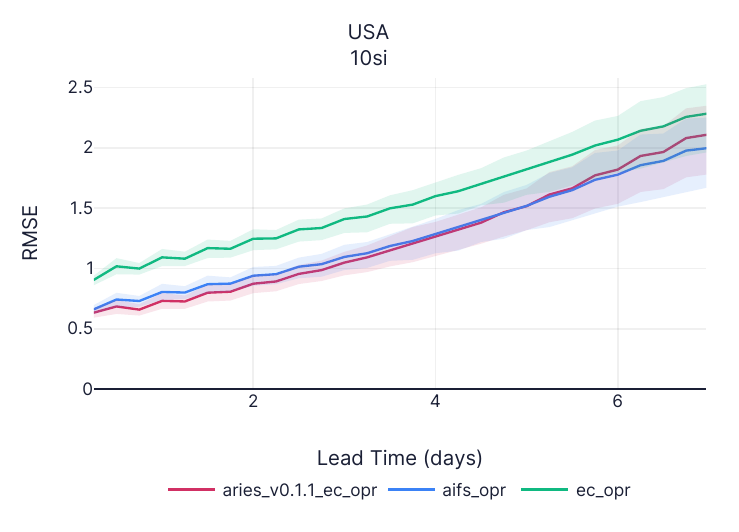}\\
        \includegraphics[width=\textwidth,trim=0 30 0 0,clip]{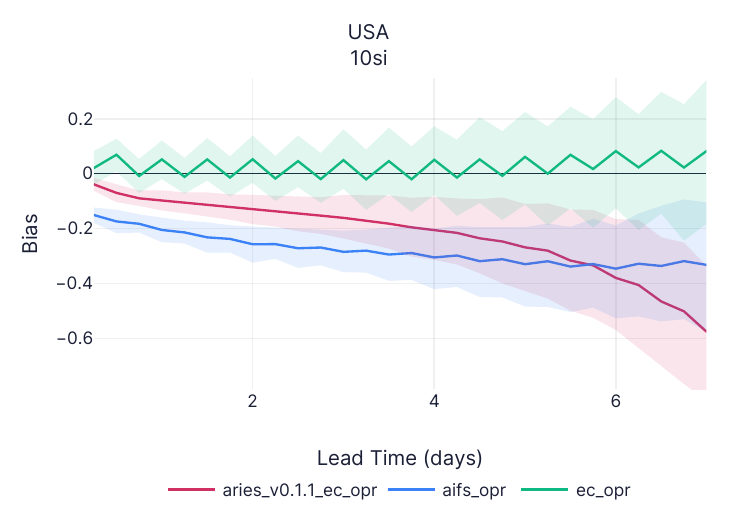}
    \end{minipage}
    \hfill
    \begin{minipage}[t]{0.32\textwidth}
        \centering
        \includegraphics[width=\textwidth,trim=0 30 0 0,clip]{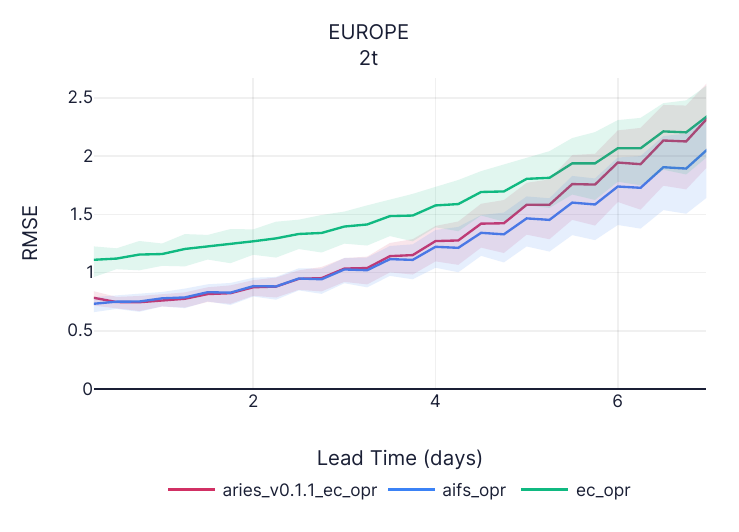}\\
        \includegraphics[width=\textwidth,trim=0 30 0 0,clip]{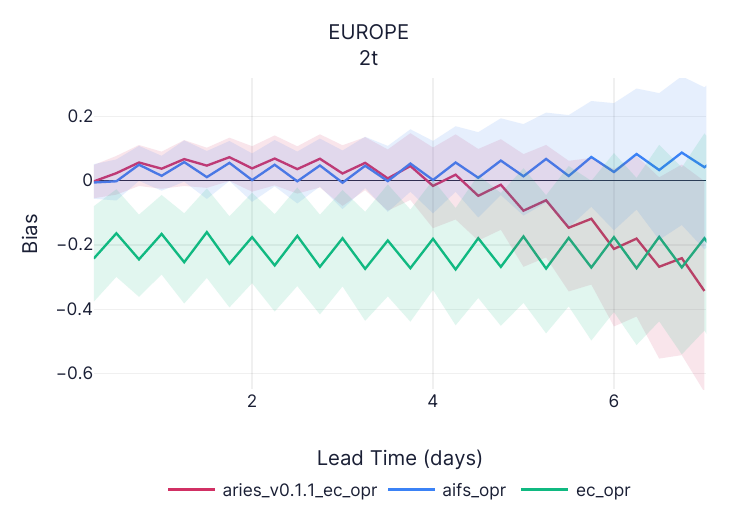}
    \end{minipage}
    \hfill
    \begin{minipage}[t]{0.32\textwidth}
        \centering
        \includegraphics[width=\textwidth,trim=0 30 0 0,clip]{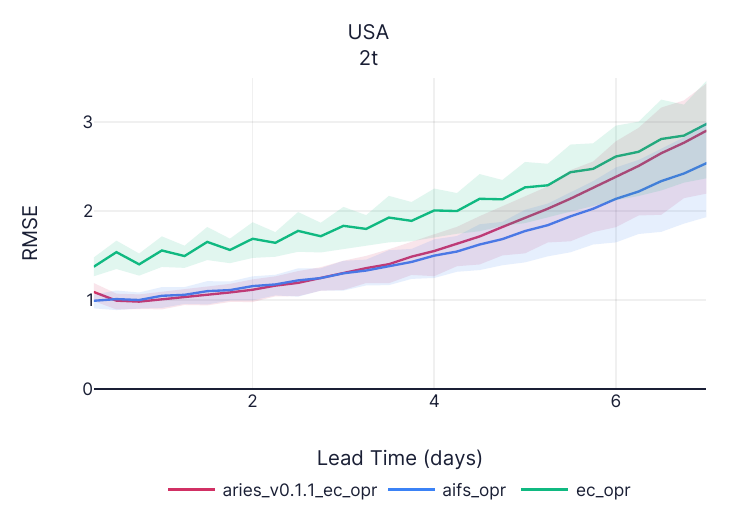}\\
        \includegraphics[width=\textwidth,trim=0 30 0 0,clip]{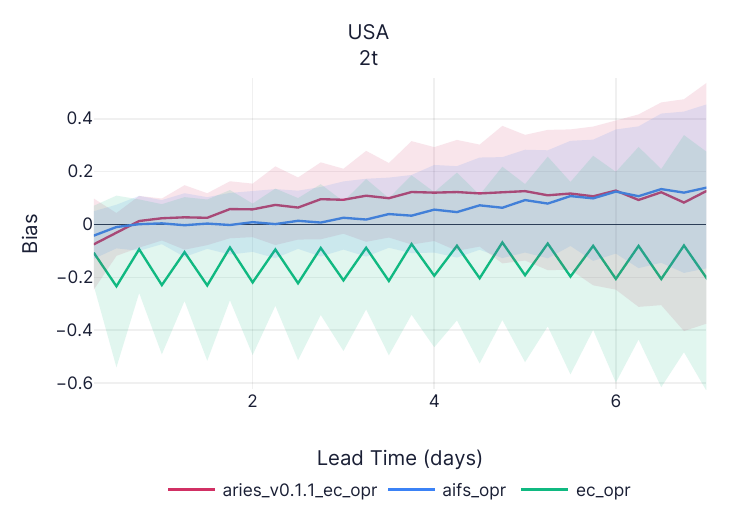}
    \end{minipage}
    \caption{Latitude-weighted RMSE (top) and bias (bottom).
    Left: 10\,m wind speed, USA.
    Middle: 2\,m temperature, Europe.
    Right: 2\,m temperature, USA.
    Shaded regions denote 25th--75th percentile ranges.
    {\color{ariescolor}\textbf{---}}~Aries,\;
    {\color{aifscolor}\textbf{---}}~ECMWF AIFS,\;
    {\color{eccolor}\textbf{---}}~ECMWF HRES.}
    \label{fig:10si_usa}
    \label{fig:rmse_2t}
\end{figure*}

\begin{figure}[H]
    \centering
    \includegraphics[width=\textwidth]{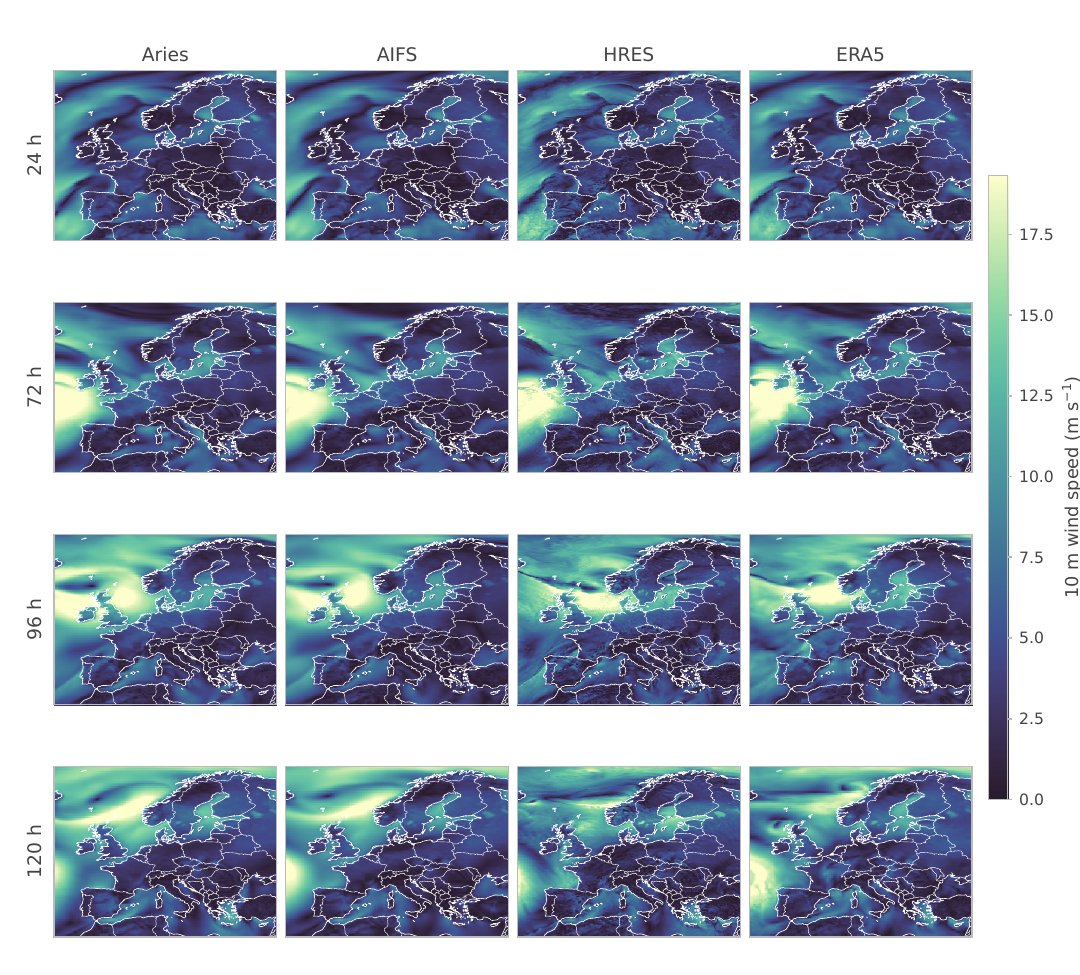}
    \caption{10\,m wind speed forecasts over Europe at 24, 72, 96, and 120\,h lead time (rows) for Aries, AIFS, HRES, and ERA5 (columns).
    The 72 and 96\,h panels capture Storm \'Eowyn over the British Isles and the North Sea.
    ERA5 is the ground-truth reanalysis.}
    \label{fig:supp_eu_10si}
\end{figure}

\begin{figure}[H]
    \centering
    \includegraphics[width=0.9\textwidth]{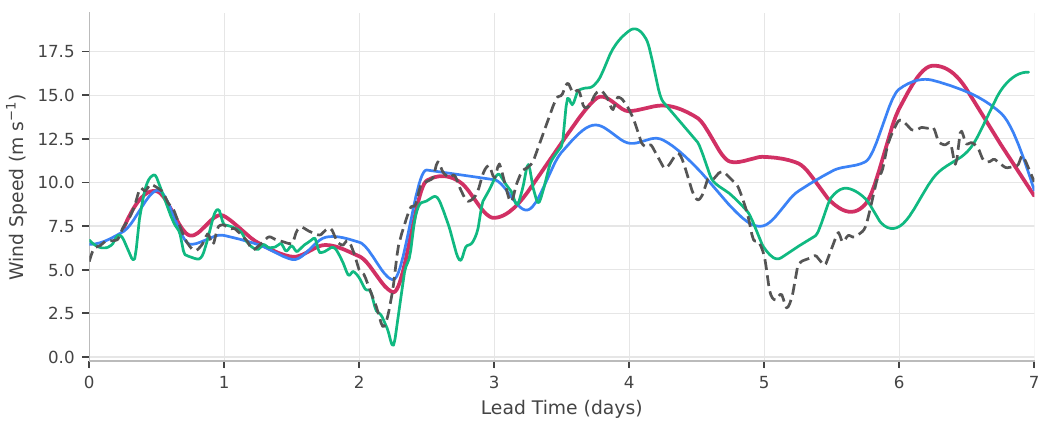}
    \caption{Time series of 10\,m wind speed forecasts at the Horns Rev offshore wind farm, Denmark ($55.5^\circ$N, $7.8^\circ$E) over the 0--168\,h horizon.
    The wind maximum around day~3.5 corresponds to the passage of Storm \'Eowyn.
    Point values are bilinearly interpolated from each model's native grid to the site; Aries and AIFS are output at 6-hourly steps, HRES and ERA5 at hourly steps, and the curves are interpolated between steps with an Akima spline.
    {\color{ariescolor}\textbf{---}}~Aries,\;
    {\color{aifscolor}\textbf{---}}~AIFS,\;
    {\color{eccolor}\textbf{---}}~HRES,\;
    \textbf{-\,-}~ERA5 (ground truth).}
    \label{fig:supp_ts_10si}
\end{figure}

Figures~\ref{fig:rmse_10si_eu} and \ref{fig:10si_usa} show the latitude-weighted RMSE of 10\,m wind speed forecasts out to 168\,h (7~days). Note that the "zig-zag" patterns are an artifact of daily error-patterns, where the two daily periods reflect the temporal shift between the 00 and 12 model runs.
Aries achieves lower RMSE than both HRES and AIFS up to approximately 96\,h (4~days) in both Europe and the USA.
This horizon aligns with the time frame over which forecasts of weather-driven supply and demand are most relevant for short-term market clearing and system operation.
In liberalized power systems, market outcomes drive generation schedules and interconnector flows.
Consequently, improved forecast accuracy in this range has the potential to enhance dispatch efficiency and system balancing.
The four-day crossover reflects the autoregressive training depth of 10 steps (60\,h); competitive performance extends to roughly 96\,h before error growth accelerates relative to AIFS.
Bias profiles (Fig.~\ref{fig:10si_usa}, right) show excellent calibration for HRES; initially small but increasing negative bias for Aries; and significant negative bias for AIFS.
Qualitative global maps and a storm case study (Storm \'Eowyn, initialized 2025-01-21) are shown in the Supplementary Material (Figs.~\ref{fig:supp_global_10si}, \ref{fig:supp_eu_10si}, and~\ref{fig:supp_ts_10si}).

\subsection{2m Temperature}

\begin{figure}[H]
    \centering
    \includegraphics[width=\textwidth]{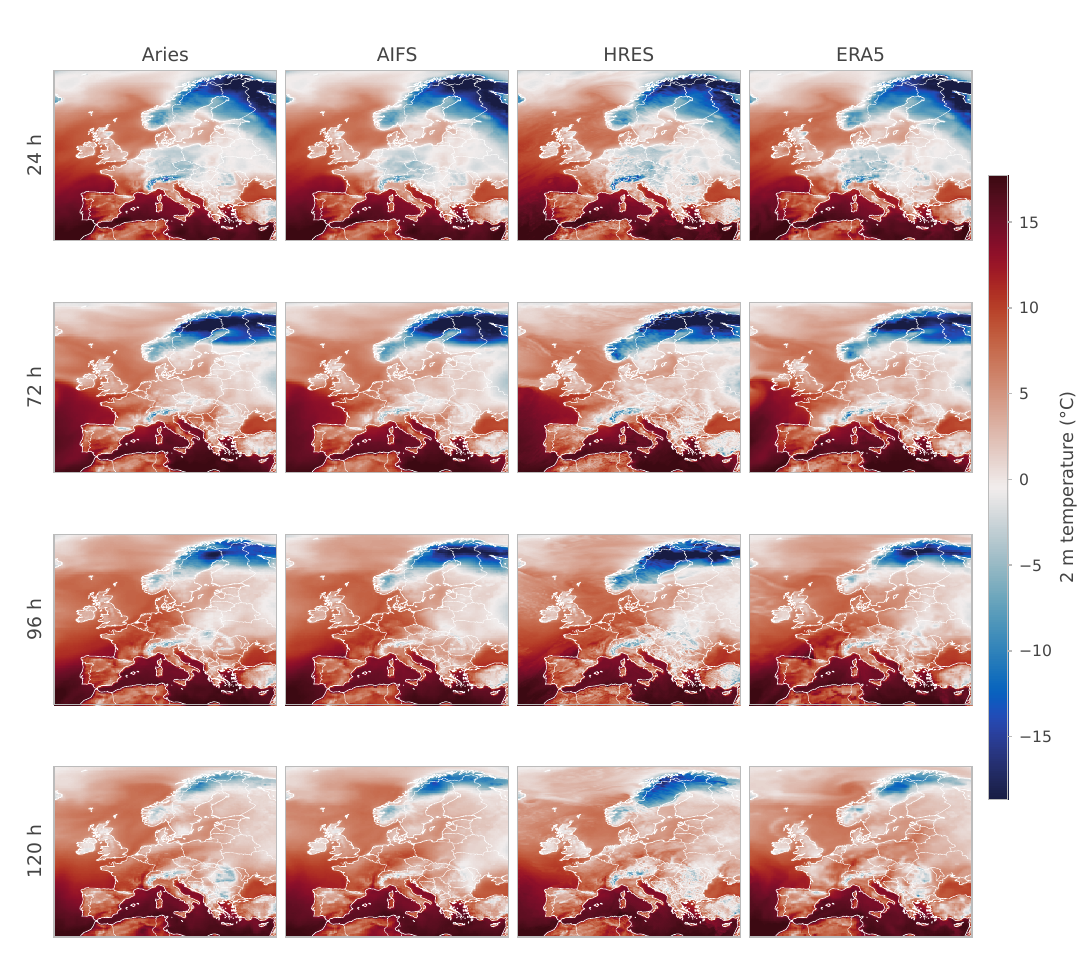}
    \caption{2\,m temperature forecasts over Europe at 24, 72, 96, and 120\,h lead time (rows) for Aries, AIFS, HRES, and ERA5 (columns).
    ERA5 is the ground-truth reanalysis.}
    \label{fig:supp_eu_2t}
\end{figure}

\begin{figure}[H]
    \centering
    \includegraphics[width=0.9\textwidth]{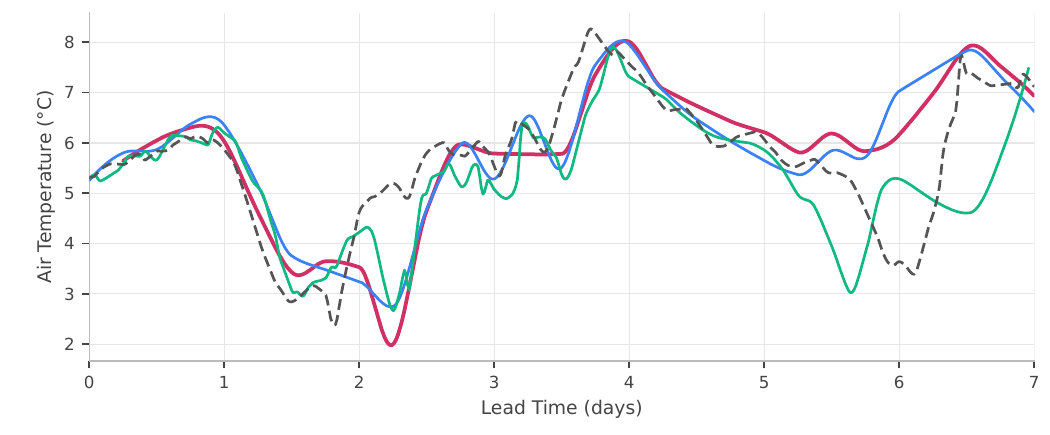}
    \caption{Time series of 2\,m temperature forecasts at the Horns Rev offshore wind farm, Denmark ($55.5^\circ$N, $7.8^\circ$E) over the 0--168\,h horizon.
    Point values are bilinearly interpolated from each model's native grid to the site; Aries and AIFS are output at 6-hourly steps, HRES and ERA5 at hourly steps, and the curves are interpolated between steps with an Akima spline.
    {\color{ariescolor}\textbf{---}}~Aries,\;
    {\color{aifscolor}\textbf{---}}~AIFS,\;
    {\color{eccolor}\textbf{---}}~HRES,\;
    \textbf{-\,-}~ERA5 (ground truth).}
    \label{fig:supp_ts_2t}
\end{figure}

Figure~\ref{fig:rmse_2t} presents the latitude-weighted RMSE for 2\,m temperature.
Aries achieves RMSE on par with AIFS during the first 96\,h, with both ML systems generally outperforming HRES against ERA5. As shown in the bias plots on the right-hand side of Fig.~\ref{fig:rmse_2t}, the RMSE discrepancy can be largely attributed to a constant negative HRES bias, due to its calibration to ECMWF Analysis rather than ERA5.
Corresponding qualitative snapshots for 2\,m temperature are provided in the Supplementary Material (Figs.~\ref{fig:supp_global_2t}, \ref{fig:supp_eu_2t}, and~\ref{fig:supp_ts_2t}).

\section{Limitations and Future Work}

Several directions remain for future work.
Verification is performed against ERA5 reanalysis only and over a single calendar year (2025) of operational runs; station-level verification would strengthen confidence in the reported skill.
The four-day crossover on 10-metre wind speed reflects an autoregressive training depth of ten roll-out steps (60\,h); extending this training horizon would push skill further into the medium range.
Aries also predicts variables of direct relevance to energy applications such as surface solar radiation downwards (SSRD) and total precipitation (TP), which warrant similar evaluation.

\section{Outlook}

As renewable capacity expands and extreme weather becomes more frequent, the value of accurate, independent forecasting compounds.
Ultimately, this benefits a broad range of entities that play an active role in modern energy systems.
More accurate short-term forecasts of wind, solar, and temperature can improve generation scheduling and reduce balancing volumes and associated system costs.
Energy trading aids this scheduling by clearing the market at a price that balances supply and demand.


We introduced Aries, a SwinTransformer-based machine-learned weather prediction model developed in industry settings.
Aries outperforms both ECMWF HRES and AIFS on 10-metre wind speed in terms of RMSE for lead times up to four days.
On 2-metre temperature, Aries achieves RMSE on par with AIFS.

We believe that diverse, independently developed weather models improve social welfare for energy suppliers and consumers alike.
Aries represents a step toward that goal, and we view the broader growth of the MLWP ecosystem as a positive development for the energy industry.


\bibliographystyle{IEEEtran}
\bibliography{references}

\onecolumn
\raggedbottom
\section{Supplementary Material}

This section presents qualitative forecast snapshots that complement the aggregate skill metrics of the main text.
All examples use the forecast initialized on 2025-01-21 at 00\,UTC, which spans the passage of Storm \'Eowyn, a record-breaking windstorm that struck Ireland, the United Kingdom, and the North Sea around 24 January 2025.

\begin{figure}[H]
    \centering
    \includegraphics[width=0.92\textwidth]{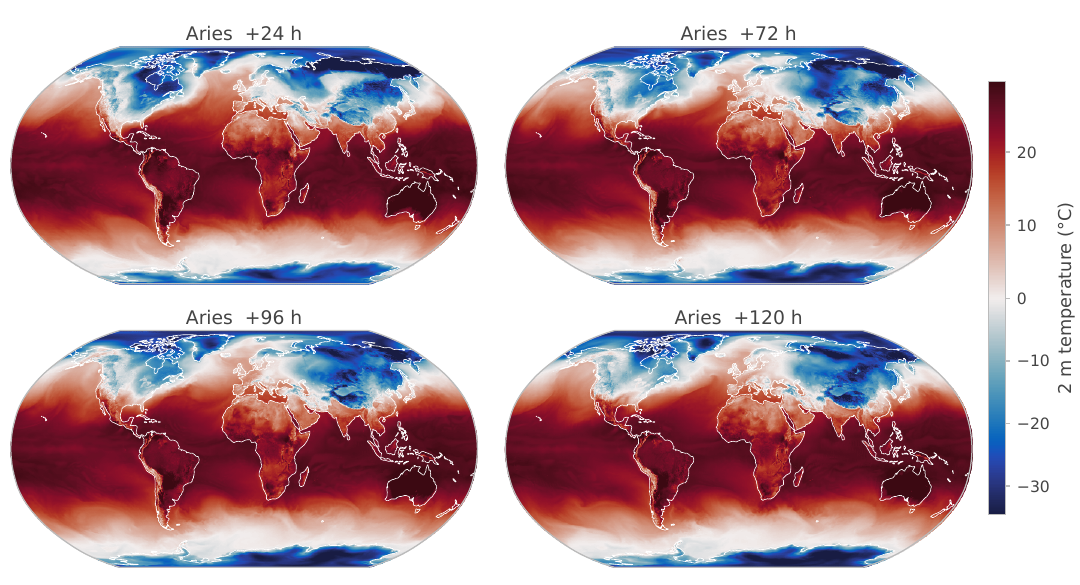}
    \caption{Global Aries 2\,m temperature forecasts at 24, 72, 96, and 120\,h lead time.}
    \label{fig:supp_global_2t}
\end{figure}

\begin{figure}[H]
    \centering
    \includegraphics[width=0.92\textwidth]{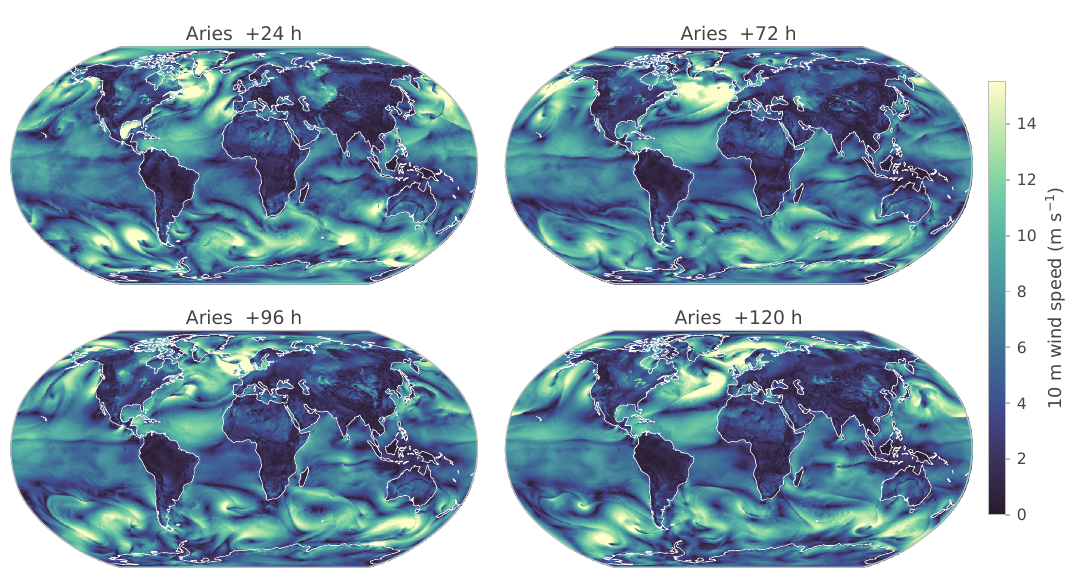}
    \caption{Global Aries 10\,m wind speed forecasts at 24, 72, 96, and 120\,h lead time.}
    \label{fig:supp_global_10si}
\end{figure}

\end{document}